\documentstyle[psfig]{mn2e}
\def\ltsima{$\; \buildrel < \over \sim \;$}
\def\simlt{\lower.5ex\hbox{\ltsima}}   
\def\gtsima{$\; \buildrel > \over \sim \;$}
\def\simgt{\lower.5ex\hbox{\gtsima}}

\input{epsf}

\title{Tidal debris of dwarf spheroidals as a probe of structure formation models}

\author[Mayer et al.]
{Lucio Mayer $^1$, Ben Moore $^2$, Thomas Quinn $^1$, Fabio
Governato $^3$, Joachim Stadel, $^4$ \\
$^1$  Department of Astronomy, University of Washington,
Seattle, WA 98195, USA mayer@astro.washington.edu, trq@astro.washington.edu\\
$^2$Department of Physics, University of Durham, South Road, DH1 3LE, Durham, 
UK, Ben.Moore@durham.ac.uk\\
$^3$ Osservatorio Astronomico di Brera,
via Bianchi 46, I--23807 Merate (LC) - Italy, fabio@merate.mi.astro.it\\
$^4$ University of Victoria, Department of Physics and Astronomy, 3800 Finnerty
Road, Elliot Building, Victoria, BC V8W 3PG Canada,\\
stadel@phys.uvic.ca}

\date{\today}

\begin{document}

\maketitle

\begin{abstract}

Recent observations suggest that Carina and other nearby dwarf spheroidal
galaxies (dSphs) are surrounded by unbound stars tidally stripped by the Milky
Way.  We run high-resolution N-Body simulations of dwarf galaxies orbiting
within the Milky Way halo to determine if such observations can be explained with dark matter potentials as those implied by current structure formation models.
We show that tidal forces acting on dwarfs with constant density cores or
with cuspy profiles having a low concentration parameter ($c < 5$) 
lead to flat outer stellar density profiles like that of Carina
for a variety of orbital configurations.
On the contrary, it is more difficult to
remove stars from cuspy dark matter halos with concentrations as high
as predicted by CDM models at the mass scale of dwarf galaxies ($c \simgt 10$)
and the data can only be reproduced assuming nearly radial orbits.  
Our simulations
show that Carina is losing mass at a fractional rate $< 0.1$ Gyr$^{-1}$ and its
mass-to-light ratio could be inflated by at most a factor of 2 due to unbound
stars projected along the line of sight.
We follow the evolution of the tidal
debris within a triaxial clumpy cold dark matter Milky Way halo which 
causes differential precession and small scale heating of the stellar streams. This renders their
use as a dynamical tracer of the Galactic potential practically useless, but
does provide a novel test of the nature of the dark matter.  Models with
warm dark matter (WDM) or collisional fluid dark matter (FDM) produce dwarf
halos with lower central densities than CDM and would be consistent with 
the observed tidal tails even for orbits with eccentricities as low 
as indicated by current data on nearby dwarf spheroidals. 
Galactic halos in FDM are
smooth and spherical and would be favored by the detection of cold
coherent streams such as that associated with the Sagittarius dwarf spheroidal.

\end{abstract}


\section{Introduction}

In hierarchical models of structure formation the first dark matter 
clumps that eventually host luminous galaxies 
have masses comparable to
the smallest dwarf galaxies in the Local Group (Haiman, Thoul \& Loeb 1996,
Haiman, Rees \& Loeb 1996,
Tegmark et al. 1997). In these models dwarf galaxies
represent the cornerstones of galaxy formation; observations of dwarf galaxies
should thus provide fundamental tests to our current understanding of
how structure formation has proceeded in the Universe.
Some of the properties of dwarf galaxy populations in the Local Group and other
nearby groups (Grebel et al. 2001) can be understood within the 
hierarchical clustering scenario; the morphology-density relation, 
namely the fact that dwarf spheroidals (dSphs)
are clustered around the dominant galaxies in the groups while
dIrrs are found at much larger distances from them 
(Mateo 1998; Grebel 2000) likely reflects the
continuous transformation of dIrrs into dSphs 
as they fall into the overdensity of the group and are
stirred by the tidal field of the primary galaxies (Mayer et al. 2001a,b).
On the other hand, the currently popular incarnations of hierarchical scenarios,
namely cold dark matter models, are challenged by
the apparent dearth of small satellites below $V_c=30$ km/s (Moore et al. 1999;
Klypin et al. 1999) and by the rotation curves of dIrrs and LSB galaxies,
that suggest that their halos have constant density cores instead of the
predicted steep cusps 
(Cote et al. 1997; Lake \& Skillmann 1991; de Blok \& McGaugh
1997, de Blok et al. 2001; McGaugh, Rubin \& de Blok 2001; Van den Bosch et al. 2001) 
These problems have recently motivated the exploration
of alternative models, such as self-interacting dark matter (SIDM, 
Spergel \& Steinhardt 2000, Moore
et al. 2000, Yoshida et al. 2000a,b; Firmani et al. 2000a,b; Hogan et al.),
warm-dark matter (WDM) (Bode et al. 2001; Dalcanton \& Hogan 2001;
Avila-Reese et al. 2001)
and fluid dark matter (FDM) (Peebles \& Vilenkin 1999; Peebles 2000).

Dwarf spheroidal galaxies in the Local Group are the faintest
galaxies known in the Universe to date ($M_B > -12$) and, in the
traditional view, their large velocity dispersions reflect 
high dark matter contents (Mateo 1998).
The long-lasting tidal interaction with the Milky Way and
M31 should strip stars and dark matter from their potentials, leading
to the formation of stellar and dark matter streams
orbiting within the halo of the primaries
(Johnston et al. 1998; 
Ibata et al. 1998;
Helmi \& White 1999).
Observational evidence of streams associated with tidally stripped 
dwarfs, such as Sagittarius, is constantly accumulating 
in the stellar halo of the Milky Way (Helmi et al. 2000, 2001; 
Martinez-Delgado et al. 2000, Ibata et al. 2001; Dohm-Palmer et
al. 2001; Vivas et al. 2001) and, more recently, even around the
nearby Andromeda galaxy (Ibata et al. 2001b).
For more distant dSphs the flattening 
of the star counts in the vicinity of the nominal tidal radius
(Irwin \& Hatzdimitriou 1995) has been often interpreted as a
signature of the presence of tidal tails 
(Johnston et al. 1999). 
However, some authors have argued that such peculiar profiles
indicate that the dwarfs are not bound and thus
no dark matter would be needed to explain the large velocity dispersions
(Kuhn \& Miller 1989; Klessen \& Kroupa 1995).
Recently, observations that take advantage
of wide field photometry have confirmed the flattening of the star counts
out to larger radii in Carina (Majewski et al. 2000; Kuhn et al. 1999),
while the cases of Draco (Piatek et al. 2001) and Ursa Minor 
(Martinez-Delgado et al. 2001) are still controversial.
Majewski et al. claim
on the basis of the observed extra-tidal extension,
that Carina must have experienced a very large mass loss rate, losing more
than 90 \% of its initial mass in 10 Gyr due to stripping by the Milky Way's tides.

The size and mass of unbound structures originating from the galaxies
must be determined by both the typical densities of the latter 
(including dark matter) and 
the scale-length of their stellar components.
Very dense systems, or systems with stellar components whose
size is such that they lie well inside the tidal radius, will be harder 
to strip (Moore 1996). 
In CDM models dark matter halos have cuspy profiles with 
inner slopes $\sim r^{-1}$ or steeper 
(Dubinski \& Carlberg 1991, Navarro et al. 1997; 
Moore et al. 1999b; Ghigna et al. 2000), 
but their typical densities within the radius of the stellar component,
measured by the concentration $c=r_t/r_s$ (where $r_s$ is the characteristic
scale radius and $r_t$ is the tidal radius) can vary
considerably (Bullock et al. 2000); the higher is the concentration
the steeper is the rise of the rotation curve and hence the
local escape speed will be also higher.

It has been shown that there is enough freedom in the structural 
parameters allowed by CDM for large galaxies in the process of merging
to form massive tails as observed (
Dubinski et al. 1996; Mihos et al. 1998;
Springel \& White 1999).
However, these conclusions cannot be trivially
extended to smaller mass scales because the typical concentration of
halos (and thus their central density)
increases substantially with decreasing mass in CDM cosmogonies.
Moreover, the choice of the orbital parameters of the interacting systems
also plays a role by defining both the typical 
intensity and the time dependence of the external tidal forces;
for the dwarf satellites of the Milky Way
and M31, we have distances and some information on the
orbits themselves (Mateo 1998), which allows us to constrain the parameter
space better than in the case of more remote binary systems. 
In addition, the subsequent evolution of the material stripped from dwarf
satellites can provide useful information on the potential of the 
Milky Way and M31; infact, the orbits that the tidal debris will
follow will reflect the underlying mass distribution.

In order to explore the role of tidal effects on stellar systems
embedded in halos analogous to those forming in cold dark matter models,
we have carried out several high-resolution N-Body simulations 
of dwarf galaxies interacting with the external potential of the Milky Way.
The simulations were performed with PKDGRAV, a fast, parallel binary
treecode (Dikaiakos \& Stadel 1996; Stadel, Quinn \& Wadsley 2002).
The paper is organized as follows: in the next section we will provide
a description of the models used for the dwarf galaxies, 
section 3 will be devoted to the results of the simulation, and finally we
will discuss and summarize our findings.


\section{Initial Conditions}

Many previous studies that investigated the tidal disruption of 
dSphs (Johnston et al. 1997; Ibata et al. 1998; Klessen \& Kroupa 1998;
Helmi \& White 1999)
employed spherical King models 
to represent their mass distribution and placed them on mostly
circular orbits in the potential of the primary halo.
The structure and orbits of the galaxies were thus detached 
from the predictions of structure formation models.
We use more sophisticated models of dwarf
satellites, whose halo masses, sizes and density profiles are consistent with
those of objects already in place at $z=1$ in
hierarchical cosmologies (White \& Frenk 1991), the dSphs being at
least as old.
The initial dwarf models are placed on bound orbits
in the Milky Way halo, which is modeled by the external potential of 
an isothermal sphere with mass $M_{prim} = \sim 3 \times 10^{12} M_{\odot}$, 
circular velocity $V_{prim}=220$ km/s, virial radius $R_{prim} = 400$ kpc
and core radius of 4 kpc (see Mayer et al. 2001b for more details). The
dwarfs are rotationally supported, exponential disks embedded in 
truncated isothermal halos with a core or within 
NFW halos (Hernquist 1993; Springel
\& White 1999) and resemble observed dIrrs. Disks and halos are sampled
by $5 \times 10^{4}$ and $3.5 \times 10^{5}$ particles, respectively.
After several orbits
the dwarfs will be transformed into dSphs and will
satisfy the morphology-density relation (Mayer et al. 2001a,b).

The halos are exponentially truncated at $r_t=R_{200}$, 
where $R_{200}$ is their virial radius (i.e. the radius encompassing
an overdensity equivalent to 200 times the average density of the
Universe).
In particular, here we consider models of low-surface brightness (LSB) 
dIrrs because these are the likely progenitors of dSphs (Mayer et al. 2001a,b).
The models with truncated isothermal halos
have a constant density core with radius $r_c$ such that $r_c=0.035r_t$.
Models with NFW halos are characterized by the concentration $c=r_t/r_s$,
where $r_s$ is the characteristic scale length of the halo, namely the
radius  at which the slope of the profile changes from $r^{-3}$ to $r^{-1}$
(Navarro et al. 1996, 1997); we consider models with either $c=4$ or $c=7$.
The satellites span a range in initial circular velocities, 
from $V_c = 20$ km/s to $V_c = 75$ km/s; across such a mass range the 
typical concentration of halos in LCDM models is $c \ge 9$ (Eke et al. 2001).
The surface density of the disks is kept fixed (except in model LIs3 and in
the model described in section 4) and corresponds
to a central (B band) surface brightness $\mu_B\sim 23.5$ mag arcsec$^{-2}$ 
assuming a stellar mass-to-light ratio ${(M/L_B)}* = 2$ 
(Bottema 1996; de Blok \& McGaugh 1997). 
Models of different masses are simply rescaled using the cosmological
scaling between virial mass, virial radius and circular velocity (Mo et
al. 1998).
The details of the procedure used to assign the structural parameters of the
halo and disks of the satellites are explained in Mayer et al. (2001) and 
Mayer et al. (2001b).

\begin{figure}
\epsfxsize=10truecm
\epsfbox{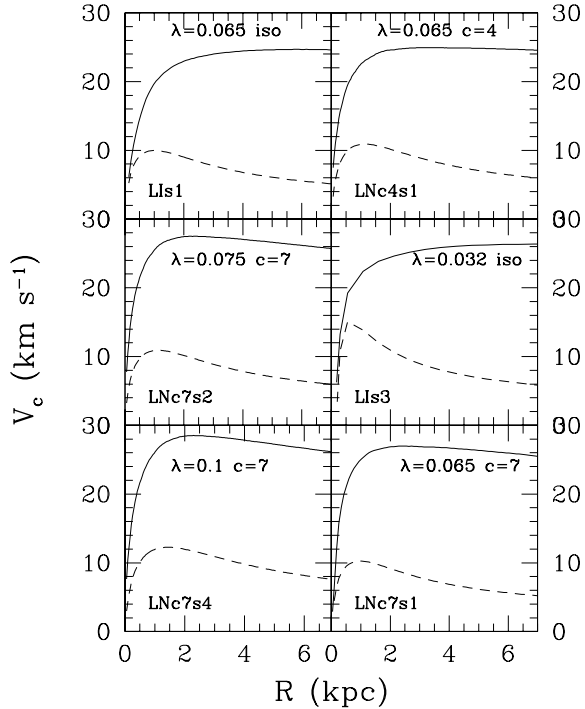}
\caption{Rotation curves of initial model galaxies for the smallest mass 
scale
considered ($V_c=25$ km/s, $M_{sat} = 3 \times 10^{-4} M_{prim}$). 
The other mass models are simply
rescaled versions of those shown here, as explained in the text. In the
panels the spin $\lambda$ and concentration $c$ (the
latter only for the satellites with
NFW halos) are indicated. We recall that the isothermal models have a fixed
core radius $r_c=0.035r_t$, while the disk scale length varies according
to $\lambda$. The related model names (used throughout the text) are in the
left bottom corner of each panel. 
}
\end{figure}

The models employed in this paper include a major improvement in that
the scale length of the disks, $r_h$, instead of being a free parameter,
is determined by the structural parameters of the dark halo, as in
current galaxy formation models. In the latter 
models it is assumed that the specific 
angular momentum of the gas, bound to the dark halo, that cools and 
eventually fall towards the center forming the stellar disk is 
initially equal to that of the halo and is conserved during the infall 
(Fall \& Efstathiou 1980; 
Mo et al. 1998).
In the simplest model in which the dark matter halo 
is a singular isothermal sphere, we
have $r_h = 1/\sqrt 2 \lambda R_{200}$, where $\lambda$ is the spin parameter, 
which measures the amount of kinetic energy stored into rotation, 
$\lambda = J {|E|}^{1/2}/ G{M}^{5/2}$ 
($E$ is the binding energy and $J$ is the total angular momentum of the halo).
N-body simulations show that the spin parameter
of large samples of halos follows a log-normal distribution, with mean values 
in the range $0.035\le \lambda \le 0.05$ 
and dispersions $\sigma_{\lambda} \sim 0.6$
(Barnes \& Efstathiou 1987; Warren et al. 1992; Gardner 2001).

Our reference isothermal LSB models (LIs1, see Figure 1) have scale-lengths
corresponding to $\lambda = 0.065$ (i.e. slightly above the average values); 
the resulting scale length is equal to the core radius of the halo, 
this being consistent with the analysis of rotation curves
in the sample of dIrrs/LSB galaxies in de Blok \& McGaugh (1997).
We also consider a model with the same $r_c$ but $r_h$ lowered by a factor
of 2  which corresponds to a smaller spin parameter, $\lambda=0.035$
(model LIs3, in Figure 1); this model has the same disk mass of
model LIs1, and thus the central surface brightness is $\mu_B \sim 22$ mag 
arcsec$^{-2}$ (1.5 magnitudes higher than the standard value).

In more realistic NFW halos the scale length of the disks depends not only on
$\lambda$ but also on the concentration $c$ and on the halo/disk mass ratio; 
the more baryons accumulate in the center, the more the whole system
contracts in response and thus the smaller is the final scale-length of the baryons.
All these aspects are taken into account in the algorithm used 
for building the galaxies (see Springel \& White 1999) but, for the
sake of simplicity, we keep the halo/disk mass ratio close to $\sim 50$ in
all our models.
The reference NFW model with $c=7$ and  $\lambda = 0.065$ (model LNc7s1 in Figure 1) 
has the same disk mass and radius of the isothermal model LIs1 but
the rotation curve rises more steeply; the rotation curve of 
the isothermal model is better reproduced
by a model with $c=4$ (model LNc4s1 in Figure 1).
We investigate also the effect of varying only $\lambda$ at fixed $c$;
models LNc7s2 and LNc7s4 have $c=7$ but $\lambda=0.075$ and 
$\lambda=0.1$, respectively (a value of $\lambda$ as high as $0.1$ is found 
in $\simlt 30 \%$ of halos in cosmological simulations).

We consider different orbital eccentricities, ranging from apo/peri=2 to 
apo/peri=15 and different disk orientations,
although we neglect cases in which the satellites are on retrograde
orbits, as it is well known that this configuration strongly inhibits
the formation of tidal tails, irrespective of the internal structure of galaxies
(Toomre \& Toomre 1972; Dubinski et al. 1996; Springel \& White 1999; 
Mayer et al. 2001a,b). The orbits have apocenters between
150 and 250 kpc and pericenters between 20 and 80 kpc, thus encompassing
the whole range of galactocentric distances of dSphs.

\section{Results}

Here we present the results of the simulations
in three separate subsections;
in 3.1 we will show that we are able to reproduce the
observed outer flat profile of Carina and that this happens
when the halo of the dwarf has a sufficiently shallow mass distribution
or when the orbit is nearly radial; in 3.2 we will analyze the
discrepancy between the intrinsic mass loss rates measured in the simulations 
and those inferred from the observations;  finally, in 3.3 we will evolve
the tidal debris of spherical dwarf models within a CDM halo
extracted from a cosmological simulation and will show how the
streams undergo precession and heating by substructure, losing
their coherence and becoming difficult to detect.

\subsection{Tidal features and the internal structure of satellites}

The satellites are severely stripped by the Milky Way tides but a bound stellar
and dark component 
survives until the end of the simulations ($\simlt 10$ Gyr); after a few
orbits (the orbital times are of the order of 2-3 Gyr), the stellar 
disk is transformed into a moderately triaxial 
system supported by velocity dispersion that closely resembles a dSph
(see also Mayer et al. 2001a,b).
Figure 2 shows the projected star counts from
four representative simulations.
In all cases we find tidal streams of stars escaping from the satellites
which show up as a flattening of the outer projected star counts.
The surface brightness of the tails depends strongly on the 
orbit and the structure of the initial models; the 
``strength'' of the streams can be quantified 
simply by the surface density of stars compared to the centrally measured 
value, $\Sigma_0/\Sigma_S$.
Satellites with cored isothermal halos or low concentration ($c=4$) NFW halos
can reproduce very closely the extended flattened star counts observed in Carina (Majewski et al. 2000), for which $\Sigma_0/\Sigma_S
\approx 10^{-2}$.  Moreover, the observations are reproduced 
even for orbits with moderate eccentricities (apo/peri=2-3) for these 
initial galaxy models. 

Instead, the streams are too weak to match the observations 
if the concentration is as high as $c\approx 7$, unless the orbits are nearly
radial (Figure 2). In fact, raising the concentration by a factor
of 2 leads to an increase of nearly 30 \% in the escape speed 
at the half mass radius of the stars where the rotation curve peaks.
(This also increases their robustness to tides since its response will
be more adiabatic, see section 3.2.)
In this case $\Sigma_S/\Sigma_0\simlt 10^{-3}$.
This is true even when the satellite is constructed with $\lambda$ considerably
larger than the mean and thus acquires a larger disk scale-length (as
in the case of the model LNc7s4 shown in Figure 2).
Infact,
the value of the concentration is the most important parameter in 
determining the amount of stripped material; a change in the spin 
affects only the disk-scale length, while a change in the concentration affects both the scale-length and the central density of the halo, or, equivalently,
its local escape speed.
However, if the initial disk scale-length is substantially decreased
while keeping the disk mass fixed, the resulting higher surface
density of the disk can affect its evolution enhancing the non-axisymmetric
instabilities that drive the morphological transformation; in fact, in
model LIs3, the high self-gravity of the compact disk  leads
to a strong bar instability and later to a very compact stellar
remnant, thereby inhibiting the developing of tidal tails (see also Mayer
et al. 2001b).

\begin{figure}
\epsfxsize=9truecm
\epsfbox{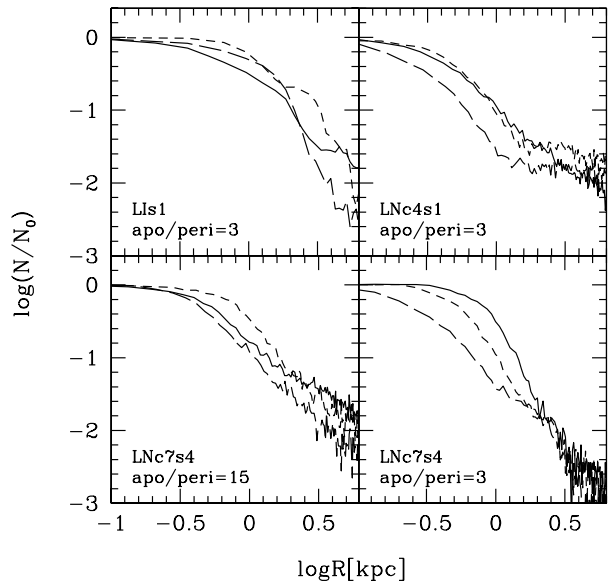}
\caption{Projected star counts (after 7 Gyr)
for representative remnants whose halos have all
the same circular velocity, $V_c=25$ km/s, corresponding to an initial 
total mass $M_{sat} \sim 10^{-4} M_{prim}$, where $M_{prim}$ is the
mass of the primary halo. 
In each panel we show the projected star counts obtained observing the stars
along one  of the tails (solid line) and along the two lines of 
sight perpendicular to the former (short dashed and long dashed lines).
The names of the models employed and the orbital eccentricities
are also indicated in the panels (the apocenter  of the orbits is
fixed at $R_{apo}=270$ kpc, so the pericenter is as small as 18 kpc
in the case of the most eccentric orbit). 
(see Figure 1 for the corresponding structural parameters).}
\end{figure}

The simulations show that the visible strength of tidal extensions is  
sensitive to projection effects. The results by Majewski et al. (2000) show
a strong flattening of the star counts at a distance comparable to
the tidal radius of the dwarf galaxy, where the counts
level out at $\sim 10^{-2}$ the peak value. The feature is more prominent
when the line of sight of the 
observer falls {\it along} one of the tidal tails, because the projected
surface density of the tails is maximally enhanced in this case. 
When viewed perpendicular to the orbit, the tails can be more
easily separated from the bound stars visually, but in this case they have the
lowest surface density (indeed corresponding to the actual intrinsic value),
$\mu_B >$ 30 mag arcsec$^2$.

In Figure 3 we show in more detail the results for the satellite 
models LIs1 and LNc4s1 with $V_c = 25$ km/s (these have an initial
total mass of only $5.8 \times 10^{8} M_{\odot}$) as these
yield remnants with physical scales close to those of Carina. 
Infact, we note that the flattening
of the star counts occurs at 1-2 kpc from the center, the
core radius of the bound remnant estimated from the fit with
a King model with $c=0.5$ is around 500 pc and the luminosity 
of the remnants is $M_B \sim - 11$, assuming a final stellar mass-to-light
ratio of  5 (as derived by combining population synthesis models with
a model for the star formation history as described in Mayer et al. (2001)).
The final dark matter mass is almost 5 times higher than the stellar mass,
yielding a central velocity dispersion $\sim 10$ km/s, as seen in Figure 4;
the corresponding final $M/L$ would be around 25 for a stellar mass-to-light
ratio of 5, in fairly good agreement with the observations (Mateo 1998)
and comfortably higher than the lower limit obtained by Lake (1990) for
galactic dwarf spheroidals.

In general, the presence of tidal extensions will lead to 
an overestimate in the velocity dispersions of the dwarf caused by
unbound stars in projection; as already
noted by others before (e.g. Piatek \& Pryor 1992), a velocity gradient or
an apparent rotation would also be observed due to the asymmetrical
contamination introduced by tidal tails.
However, in our simulations the 
apparent velocity dispersions in the central region of dwarfs are at most
$\sim 40 \%$ larger than the 
intrinsic value, which would inflate the measured $M/L$ by no more than 
a factor $\sim 2$ (giving $M/L \simlt 50$ in the remnants LNc4s1 and LIs1).
Figure 4 clearly shows that only measurements extending
a few kpcs from the center would be significantly altered; therefore
the contamination of kinematics induced by tails would not explain
the high mass-to-light ratios of dSphs without dark matter,
contrary to previous  claims (e.g. Kuhn \& Miller 1989; Kuhn 1991).

\begin{figure}
\epsfxsize=9truecm
\epsfbox{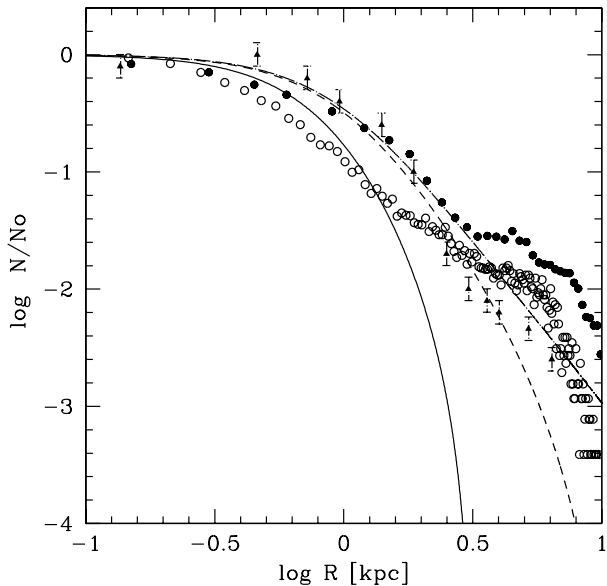}
\caption{Projected star counts obtained for the remnant of model LIs1
(filled circles) and model LNc4s1 (open circles
with $V_c=25$ km/s, both observed along one
of the tidal tails. The dots with error bars 
are the number counts from Majewski
et al. (2000) averaged over the four samples of giant stars. 
Fits with King profiles with c=0.5, 1, 1.5 
(where $c=log_{10}(r_t/r_c)$) are
also shown (solid, short dashed and long dashed lines, respectively).
}
\end{figure}

Johnston et al. (1999) found that the contamination of streams in the 
projected star counts can show up even inside the intrinsic tidal radius
of the satellites if the viewing angle is not perpendicular to the tails.
We determined the final tidal radii of our remnants with SKID, that identifies
all the particles bound to the dwarf; we found that marginally bound
particles are found as far as 3-4 kpc from the center, namely at a  
larger distance than the break radius in any of the projections. However,
this marginally bound region (that corresponds to the very inner part of the
tails) comprises only a few percent of the total mass of the remnant. 
We conclude that, although the break radius is not equivalent to the tidal
radius, it roughly defines the boundary of the remnant, encompassing 
$\simgt 90\%$ of the bound mass.

\subsection{Determining mass loss rates using extra-tidal stars}

Estimating the mass loss rate of satellites from extra-tidal features is not a
straightforward task.
Johnston et al. (1999) have proposed a formula for estimating
the mass loss rate of an orbiting stellar system which relies on the
assumption that the number counts surface density profile of the extra-tidal 
material is $\Sigma_{xt} \sim r^{-1}$.
Majewski et al. (2000) have applied such a formula to the star counts profile 
they obtain for Carina and derive an extremely large 
fractional mass loss rate, $df/dt=0.3$ Gyr$^{-1}$. 
This would imply that the mass of the stellar 
component in Carina would have been reduced by as much as a factor 
of 30 in 10 Gyr. We obtain comparable fractional mass loss rates
applying the same formula to the remnants in our simulations, while the 
actual average fractional mass loss rates measured in the
simulations (obtained by simply dividing the total mass lost 
by the duration of the simulation of about 10 Gyr) 
are significantly lower, $\sim 0.06 {\rm Gyr}^{-1}$.

\begin{figure}
\epsfxsize=8truecm
\epsfbox{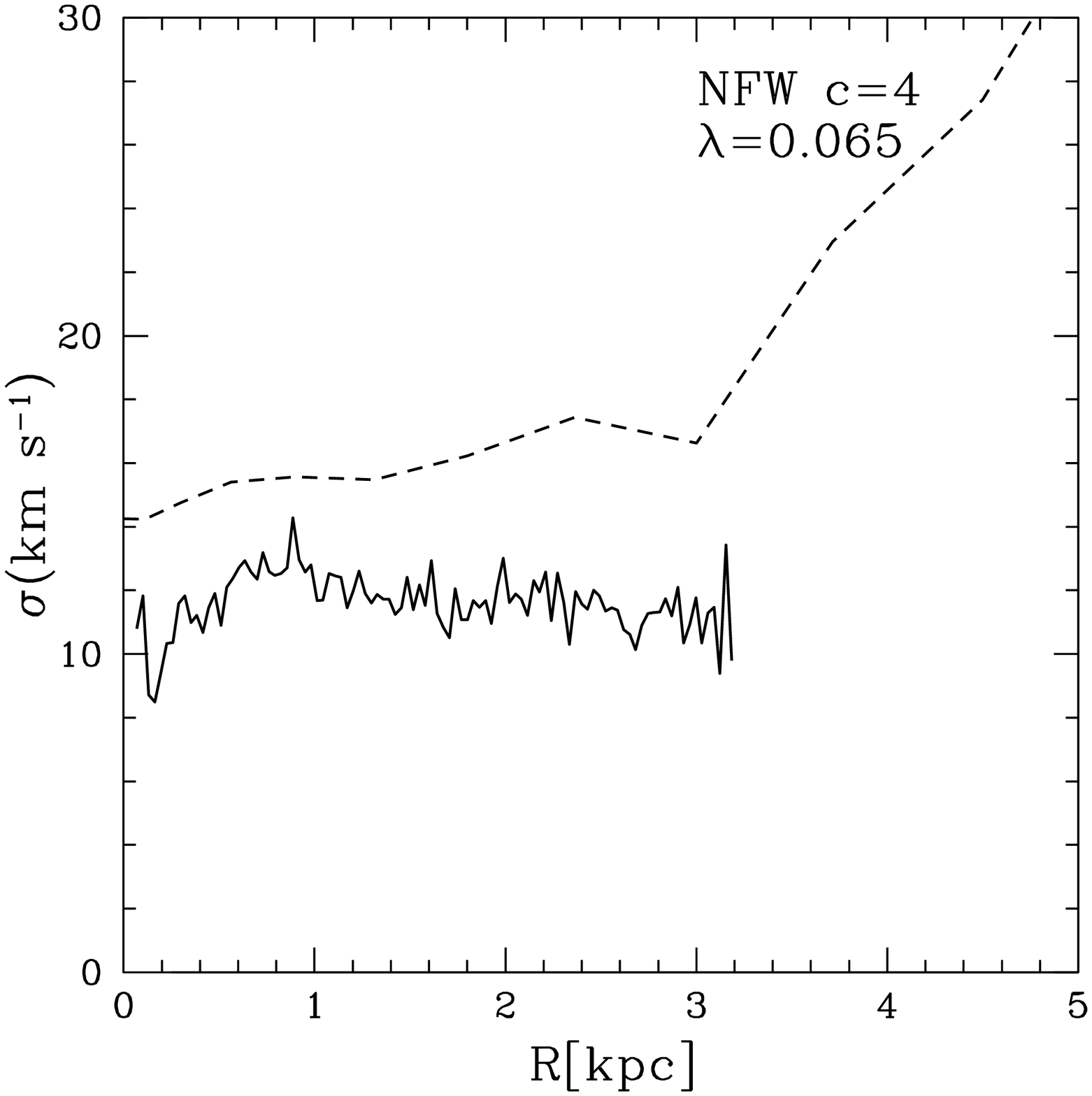}
\medskip
\epsfxsize=8truecm
\epsfbox{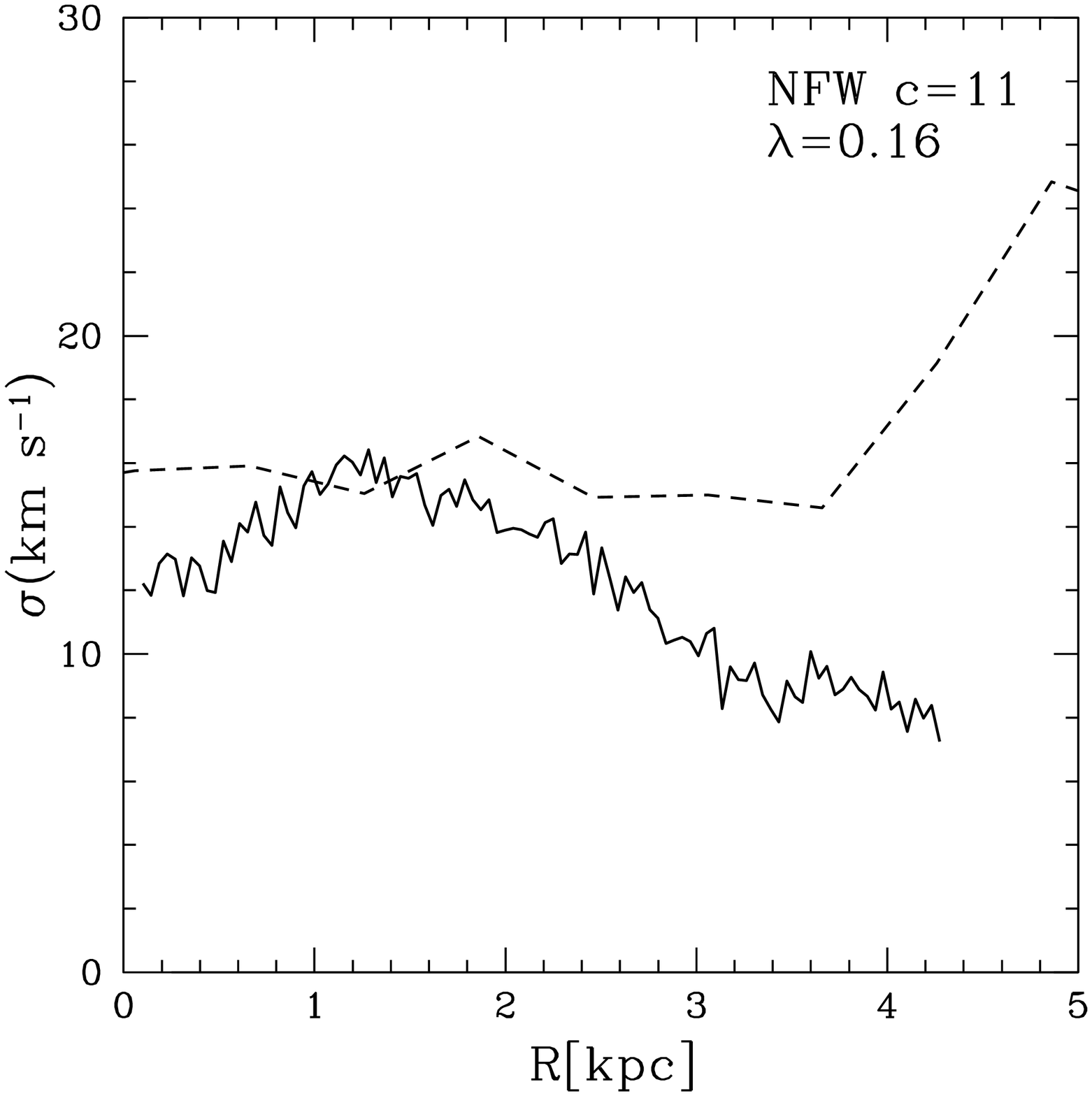}
\caption{Intrinsic (solid line) versus apparent stellar velocity dispersion 
profiles (dashed line) for a line of sight that intercepts one of the 
tidal tails. The models used in the simulations are LNc4s1 (upper panel)
and the model with a highly concentrated halo and a large disk discussed
in section 4.1. The orbits have apo/peri$=2$, with $R_p=55$ kpc and both 
have halos with an initial circular velocity $V_c = 25$ km s$^{-1}$.}
\end{figure}

One problem in the application of the formula
is the assumed radial surface density profile of the stream.
Infact, the outer star counts profile of our remnants are
much flatter than $r^{-1}$ for most of the viewing projections.
Removing the constraint on the profile of the 
extra-tidal stars, Johnston et al. (1999) derive also
an upper limit for the mass loss rate:

\begin{equation}
{df \over dt} = cos(\theta) {\Sigma_{xt}(r_{break}) \over 
n_{break}} {\pi \over T_{orb}} 2 \pi {r_{break}}^2
\end{equation}

where $\theta$ is the angle between the velocity vector of the satellite and
the line-of-sight (which can be derived for those satellites that have
measured proper motions), $n_{break}$ is the number of stars counted 
within $r_{break}$ (the radius where the profile 
changes slope) and $T_{orb}$ is the orbital time.
Using equation (1) for the cases with
clear tidal features we obtain  $df/dt < 0.5$ Gyr$^{-1}$,
which is still not representative of the numerical results, being higher
than the real mass loss rate by nearly an order of magnitude.

Equation (1) is derived under the assumption 
that the stars are lost continuously
over the orbital time $T_{orb}$. However, the tidal mass loss 
for a satellite moving on an eccentric orbit will occur mostly as
a result of the tidal shocks suffered at pericenter and 
will depend on its ability to
respond to the perturbation and adjust to a new equilibrium.
To gain a better insight into the mechanism, we can consider
a galaxy that suffers a tidal shock of duration $\tau = R_p/V_p$, where $R_p$ 
and $V_p$ are, respectively, the distance and the velocity of the galaxy at 
the pericenter of the orbit.
The characteristic time required for a star at the disk half mass radius 
$r_{hm}$ to increase its mean-square velocity $<v^2>$ due to the shock is 
$t_{shock} \sim \tau {V_c}^2/<{\Delta v}^2>$, where $V_c$ is 
the internal (rotational) velocity of the galaxy and $\Delta v$ 
is the velocity impulse; this timescale will be comparable to the 
characteristic  timescale of mass loss  (see Binney \& Tremaine, 1987).
Gnedin, Hernquist \& Ostriker (1999) have provided analytical formulas
to calculate  $\Delta v$ for a satellite that is shocked by an extended
perturber with an isothermal profile 
including even the adiabatic corrections that account for the motion
of the stars within the galaxy. As shown by the authors, if the orbits are
eccentric, a good approximation is obtained by simply assuming that the
satellite is moving on a straight path. In the
latter case we can write 

\begin{equation}
<{\Delta v}^2>={{\left (GM_0 \over{{R_p}^2 V_p}\right)}^2 {r^2 \pi^2\over 3} 
{R_p^2 \over {R_{max}}^2} \times {(1 + \omega\tau)}^{-2.5}}
\end{equation}

where $M_0$ and $R_{max}$ are the total mass and radius
of the perturber and the last term in the product is the 
first-order adiabatic correction
($\omega$ is the typical stellar frequency at $r=r_{hm}$,
and $\omega\tau = \tau/t_{dyn} \sim 1$ in our models, where $t_{dyn}$ is
the dynamical time at $r=r_{hm}$).
For a satellite with $V_c=25$ km/s and $r_{hm} = 2$ kpc on
the apo/peri=3 orbits in our simulations ($R_p =75$ kpc, $V_p \sim 300$ km/s),
we obtain $t_{shock} \sim 5$ Gyr. We can consider $t_{shock}$ as an estimate
of the  characteristic time over which the satellite loses 50 \% of its 
initial disk mass.

As we can see from Figure 5, the satellites on apo/peri=3 orbits
actually have lost about 40\% of their mass
5 Gyr after the first tidal shock. Although in the simulations 
more than one shock occurs, the non-axsymmetric instabilities have a 
counteracting effect, as they make the stellar component more concentrated 
and increase  the adiabatic corrections over time (see Mayer et al. 2001b), 
thereby explaining the good agreement with the above simple estimate.

\begin{figure}
\epsfxsize=8truecm
\epsfbox{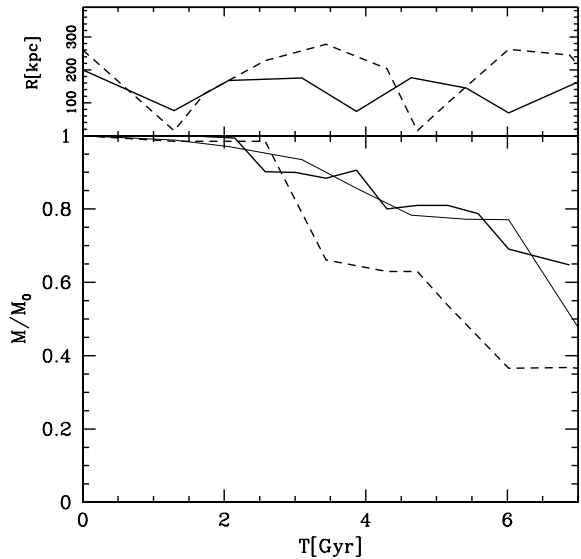}
\caption{Fractional mass loss over 7 Gyr for satellites 
on a orbit with apo/peri=3 and $R_p = 75$ kpc (the thick solid line 
refers to model LIs1 and the thin solid line to
model LNc4s1) and on a orbit with apo/peri=15 and  $R_p =$ 20 kpc (the dashed
line, corresponding to model LIs1).}
\end{figure}

The resulting average fractional mass 
loss rate is $\sim 0.1/{\rm Gyr}$. The agreement with the simulations strongly
depends on the adiabatic corrections; if we neglect them then the resulting
mass loss rate would be higher by a factor of 5 and would be
close to the predictions of Johnston et al. (1999). 
In the inner regions of the system the adiabatic corrections will be
increasingly more important and will increase substantially the 
overall lifetime of the satellite. 

Using the numerical results as a guide we propose a simple recipe to
infer the mass loss rate from the observed star counts profile;
on typical orbits as those considered here the dwarfs lose mass
at a rate that scales as $\sqrt{(\Sigma_S/\Sigma_0)}$;
employing the simulations that match the data on Carina
to calibrate our estimate 
(these yield a fractional mass 
loss rate $df/dt \sim 0.06$ Gyr$^{-1}$), we obtain 
$df/dt \sim 6 \sqrt{(\Sigma_S/\Sigma_0)}$ Gyr$^{-1}$. This result 
was derived for the satellites on orbits with apo/peri$=3$, but the
mass loss rates on nearly radial orbits differ by less than a factor
of 2 (see Figure 5).

\subsection{Precession and heating of tidal streams}

For dwarfs orbiting in spherical and smooth potentials like those considered
so far the escaping stars form symmetric 
tidal tails that lead and trail the satellite revealing its future and past
orbit (see also Moore \& Davis 1994). However, in general the orbital 
evolution of the streams will depend on the structure of the 
underlying potential and in turn the streams might be used as a tracer 
of such structure only provided that they remain suficiently coherent 
(Johnston et al. 1999a, Zhao et al. 1999).
Cold dark matter halos are
complex triaxial systems with shapes and angular momenta that
vary considerably from the center to their virial radii (Moore et al 2001).
They also contain dark matter substructure and satellite galaxies that
can heat and perturb parts of the stream away from their orbital 
paths. These effects might combine to destroy the coherent nature of tidal
streams through differential precession of orbits (see also Johnston,
Sackett \& Bullock 2001).

To investigate the long-term evolution of tidal tails we construct 
a massless spherical system using 50,000 particles distributed with a 
density profile $\rho(r)\propto r^{-1}$, radius 10 kpc and velocity 
dispersion $\sigma_{1d}=10$ km/s. This unbound test satellite will form tidal
streams that can be used to explore orbits within different halos. 
The evolution of this system on a series
of circular orbits within a smooth spherical potential is shown in Figure 6. 
This potential is a singular dark matter halo with density profile 
$\rho(r)\propto r^{-2}$ with constant 
velocity dispersion $\sigma_{halo}=200/\sqrt{2}$ km/s. The rings of debris correspond to
satellites on initially circular orbits at radii of
$0.2, 0.4, 0.6, 0.8, 1.0r_{200}$ where $r_{200}=300$ kpc.

The satellite is unbound and immediately starts to form symmetric tidal
tails because the model has a finite size and the particles have random velocities.
The tidal debris lies in the orbital plane and after a time scale of
6 Gyrs the particles lie in streams that wrap around more than one orbit at the
center of the potential and about 10\% of an orbital radius at the edge.

We now repeat the test using a CDM halo taken from one of the cosmological 
simulations of Moore et al. (1999a). By construction, 
the circular velocity and virial radius are similar to 
those adopted in the spherical potential above. The halo is resolved with $10^6$
particles within $R_{200}$ and we use a comoving softening length of 0.5 kpc.
This particular galactic mass
halo virialises by a redshift z=0.5 and does not accrete any bound object
containing more than 1\% of its mass by the present day.
We calculate the axis with the largest component of angular momentum 
and place the test satellites in circular orbits at the same radii as used 
in the spherical potential above.

\begin{figure}
\epsfxsize=8.5truecm
\epsfbox{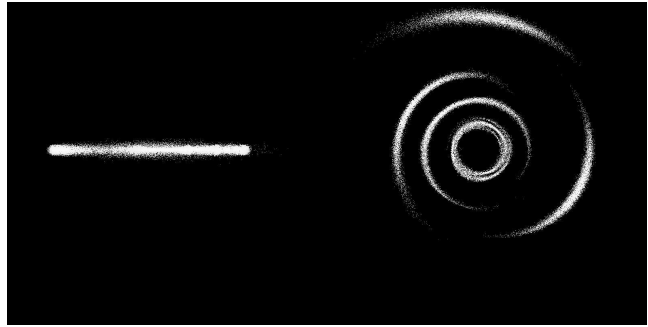}
\caption{The evolution of five massless satellites on circular orbits within
a smooth spherical potential for a period of 6 Gyr (comparable to the
orbital time of the outer satellite). The left panel is an edge on view
of the orbital plane and the right panel is a face on view.}
\end{figure}

\begin{figure}
\epsfxsize=8.5truecm
\epsfbox{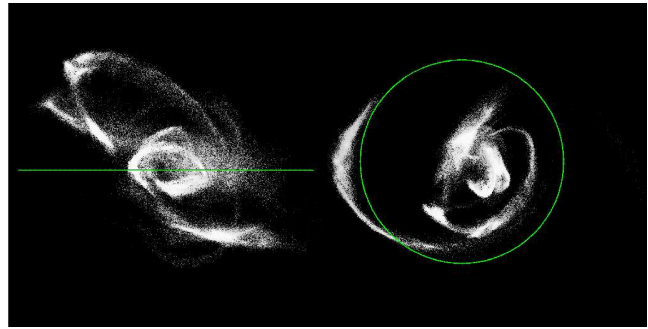}
\caption{
The evolution of 5 massless satellites on initially circular orbits
(same as in Figure 6) within a CDM galactic mass halo. 
The satellites are introduced at a
redshift z=0.5 and the simulation is continued to the present day. 
The left panel is an edge on view of the initial orbital plane and the 
right panel is a face on view.}
\end{figure}

Figure 7 shows the satellite particles after a period $\sim 6$ Gyrs revealing
that the tidal streams have precessed dramatically away from their initial orbits.
The inner streams have wrapped several times around the halo, both within and out of the
orbital plane whilst the outer streams remain more coherent but have precessed from
their initial orbital plane by over 45 degrees. The precession from the 
differentially rotating triaxial halo structure is mostly responsible for
the differences between the orbits of the tidal debris between Figure 6 and Figure 7. 
However, the substructure in the galactic CDM halo has also contributed visible
heating to the streams. This is most noticeable in the outer streams which
have a clumpy appearance due to heating by the most massive subhalos.
(Dark matter substructure in the outer halo will be more massive and 
cause the most heating.). 
Numerical heating of the streams due to
massive halo particles (Moore et al. 1996) might somewhat enhance the observed 
evolution of the streams. 
In order to test our results we ran  a  massless spherical
satellite in a $10^5$ particles Milky Way-sized halo with an 
NFW profile; this halo is similar to the CDM halo as far as
profile, mass and radius are concerned but it is spherical
and does not have substructure, therefore no physical heating or precession
is expected. After 10 Gyr we measured fluctuations of only a few percent
in the orbital energy and orbital angular momentum, and, correspondingly,
the initial orbital plane of the satellite is barely altered; this shows that 
two-body heating has a negligible contribution to the evolution
of the streams (this is a conservative test since the CDM halo in the
cosmological volume had a resolution almost ten times higher).

The experiment illustrated in this section
ignores the dissipative effects of a baryonic component that would
help regularise the structure of the inner dark matter halo. However, the outer
halo would remain unaffected by the adiabatic contraction and this
tends to produce oblate inner halos (Dubinski \& Carlberg 1991). 
Coherent streams should only be found in the plane of the disk or on
polar orbits. Our results suggest that using streams to constrain the
potential structure of galactic halos will be complicated by these effects.
Evidence of streams on great circles in galaxy and cluster halos would
support models in which the dark matter behaves 
like a fluid. In this case, halos are highly spherical and contain less mass
attached to subhalos due to ram-pressure stripping.
Interestingly, the stream
associated with the Sagittarius dwarf spheroidal appears as a great circle;
Ibata et. (2001), based on the analysis of the latter, have recently 
argued that the  Galactic halo must be nearly spherical between 16 and 60 
kpc from the center.

\section{Discussion}

We have shown that the outer flattening observed in the stellar
profiles of some dSphs can provide a useful constraint on the
structure of their dark matter halos once their orbital eccentricity is known.
At present, the few proper motions available (e.g. Kroupa \& Bastian
1997 for the LMC and SMC, Schweitzer et al. 1995 for Sculptor and
Schweitzer \& Cudworth 1996 for Ursa Minor) imply
that the Galactic satellites 
have orbits with low eccentricities, with typical apo/peri=2-3.
It is clear that systems on these orbits with concentrations
as high as predicted in the standard LCDM cosmology (Eke et al. 2000)
for galaxies with $V_c < 50$ km/s ($c \ge 9$) would not be able to 
reproduce the observed star counts of the Carina dSph.


One might argue that the properties of halos in CDM models
have a large scatter. However, recently Bullock et al. (2001) have measured
the scatter of the concentration of halo profiles in LCDM simulations and,
according to their results, a concentration of $c=7$ is already a lower limit
for halos with $M \sim 10^{11} M_{\odot}$ (i.e. corresponding to the most
massive among the models considered in this paper), suggesting that 
we are being conservative with our constraints.

Satellites with either softened isothermal halos
or cuspy halos with low concentrations, exhibit features that 
can reproduce the observations -- 
even on orbits with moderate eccentricities. 
Interestingly we note that the values of the concentration needed
($c < 5$) are as low as those found within WDM models at the dwarf galaxy scale
(Eke et al. 2001).

\subsection{High mass to light ratios in dSphs}

Our results have implications on the actual 
dark matter content of dSphs. Large tidal tails are produced only when
the dark matter in the remnants accounts for $ \le 80$ \% of the 
total mass. The apparent $M/L$ can be at most two times higher
than the intrinsic value because of the unbound stars projected 
along the line of sight and therefore we expect to find tidal tails 
only around dwarfs whose measured $M/L$ is not very high.
Indeed most of the dSphs in the Local Group, including Carina, 
have $M/L \le 30$ (Mateo 1998).

The dark matter contents of Draco and Ursa Minor, having $M/L > 60$
(Hargreaves et al. 1994a.b, Mateo 1998)
are instead too high even accounting for enhanced apparent velocity
dispersions. Such high dark matter contents would be
incompatible with the observations of massive tails; the remnants that
have similarly large dark matter contents at the end of our simulations 
(see also Mayer et al. (2001a,b) on the evolution of the GR8 model)
have star counts at the level $\Sigma_{S}/\Sigma_0< 10^{-3}$ in the outer 
part. 

Although very recent results exclude the presence of tails 
in Draco (Odenkirchen et al. 2001; Aparicio et al. 2001), other authors
have reported positive detections of some extra tidal stars around the
latter dwarf and Ursa Minor (Piatek et al. 2001,  Martin\`ez-Delgado 
et al. (2001)). In order to further explore this issue we ran a simulation with
a satellite model different from those used so far. We used an NFW
halo with a high concentration, $c=11$, and $V_c = 25$ km/s, and a 
disk extending out to half of the virial radius of the halo, $\approx 10$ kpc,
corresponding to a high spin parameter, $\lambda =0.16$, and a very low
central surface brightness, $\mu_B \sim 25$ mag arcsec$^{-2}$.

We placed the model on an orbit with an apo/peri$=2$ and a pericenter $R_p
=55$ kpc, consistent with the present distances and radial velocities of Draco
and Ursa Minor (Mateo 1998). After 7 Gyr the satellite has turned
into a dSph ($v_{rot}/\sigma \sim 0.3$)
with a central velocity dispersion $\sigma \sim 10$ km/s
and a dark matter halo still 7 times more massive than the stellar
component; this would yield $M/L > 40$ for a stellar
mass-to light ratio $\sim 5$ (Mayer et al. 2001b). Furthermore,
due to its unusually large scale length, the stellar component has undergone
severe stripping, producing prominent tidal tails and signatures
in the star count profile qualitatively similar to
those observed in Carina (Figure 8); including possible projection
effects due to the tidal tails, the apparent 
mass-to-light ratio could be enhanced up to values around 80. 
This experiment shows that
an extremely extended disk can produce
large tails notwithstanding the high central dark matter density of the halo.
However, a closer look shows that the remnant is very extended in radius
and the  flattening of the star counts in
the remnants occurs at more than 4 kpc, which is too far out compared
to the size of Draco, the latter being less than 1 kpc (Mateo 1998).
In addition, a dwarf progenitor with an extremely extended disk is hard
to support from the observational point of view; 
the rotation curves of dIrrs and LSB spirals suggest that 
the scale-lengths of halos and baryons are correlated 
(de Blok \& McGaugh 1997) and in some cases the scale-length of the halo
can be larger than the typical disk size (Lake \& Skillmann
1990), while in the present model is considerably smaller.
in the present model would clearly cause 
Besides, the surface density of the gas in the outer part of
such an extended disk would be so low that the star formation would have
hardly occurred, making the formation of stellar tails unlikely.

\begin{figure}
\epsfxsize=8truecm
\epsfbox{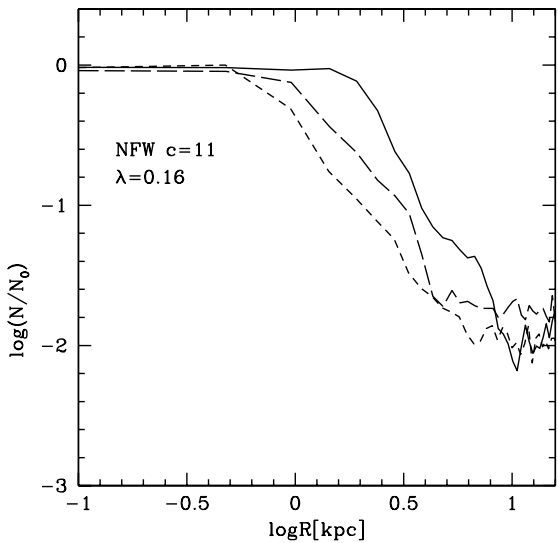}
\caption{Projected star counts for the remnant (after 7 Gyr) of a satellite 
model with a highly concentrated halo. The satellite has $V_c=25$ km 
s$^{-1}$ and was placed on a orbit with apo/peri$=2$ and $R_p=55$ kpc.
We show the projected star counts along one 
of the tails (solid line) and along the two lines of sight perpendicular to the
former (short dashed and long dashed lines).}
\end{figure}

The correct evaluation of the $M/L$ of dSphs is important when one tries to
compare the observed number of Galactic satellites with the numbers predicted
by cosmological N-body simulations (Moore et al. 1999; Klypin et al. 1999).
It has been pointed out that dwarfs with NFW halos
should have rising velocity dispersion profiles and that by using
the measured {\it central} velocity dispersions one would actually 
underestimate the total mass of dSphs (White 200); 
including this correction would improve the agreement with observations
(Moore 2001).

Our simulations suggest two possible scenarios 
with regard to this issue and both are highlighted in Figure 4.
First, if it were confirmed that most of the satellites
have large tidal tails and that their orbits have low eccentricity, then
it is very likely that their halos have low concentrations, $c < 7$,
and have quite flat intrinsic and apparent velocity dispersion profiles; 
the apparent velocity dispersion can actually be higher than
any value of the intrinsic velocity dispersion, but never smaller, which
would  actually  tend to {\it reverse} the correction, (cfr. Figure 4a).
Secondly, even if the satellites had more concentrated NFW halos
(1) the intrinsic peak velocity dispersion is only $\sim 25 \%$ higher
than the central value in the remnants and (2)
the intervening tidal debris, though quite weak, tend 
to flatten even further the apparent velocity dispersion profile
(cfr. Figure 4b). 
Hence, it seems that a better assessment of the actual
mass of the dwarfs derived from apparent velocity dispersions
would never solve the substructure problem and could 
actually make it worse.


\section{Summary}

We have used numerical simulations of two component models of dwarf galaxies
orbiting within a Milky Way potential to study the effects related
to their tidal tails. The signatures produced
by the tidal debris in the stellar profiles 
can be used to constrain the structure of the dark matter halos of dwarfs,
providing hints to the nature of the dark matter.
The long term evolution of tidal debris was followed 
within a smooth spherical potential and within a high resolution cold
dark matter halo. We summarize our conclusions here:

${\bullet}$ We can produce tails as prominent as observed 
around the Carina dSph for models with soft central potentials
or singular potentials on highly radial orbits. Knowledge of the orbits of the dSphs 
is vital to constrain directly their internal structure. If it
will be confirmed that the orbits are nearly circular, warm
dark matter models will be favored instead of cold dark matter.

${\bullet}$ Models with very high mass to light ratios and strong tidal
tails are very difficult to produce. Draco and Ursa Minor should only have
very weak tails, at a star count level at least ten times 
lower than that found in Carina.

${\bullet}$ Our simulations provide a simple estimate of the mass
loss rates of dSphs based on star counts and these are significantly
lower than those obtained by others discarding adiabatic corrections;
Carina is likely to survive for at least another 5 Gyr.

${\bullet}$ Only satellites
on very tightly bound orbits like Sagittarius, that suffered many
strong tidal shocks, could have been already destroyed in the past or 
could be now close to disruption.

${\bullet}$ Central velocity dispersions of stars in dSphs are good indicators
of the characteristic peak 
velocity dispersions of their dark matter halos. i.e. tidal
heating produces nearly isotropic orbits and quite flat velocity dispersion
profiles.

${\bullet}$ The masses of dwarf spheroidals could
be overestimated when large tidal tails are present but the maximum effect due
to projection of unbound stars is only a factor of two. However, this
would shift the mass function of galactic satellites further away from that 
predicted  by cold dark matter models, making the substructure problem 
even worse.

${\bullet}$ The tidal debris within a cold dark matter halo shows dramatic 
signatures of differential precession and some heating by 
dark matter substructures. It may be impossible to use observations
of tidal streams to map the structure of the Galactic halo.

${\bullet}$ If many examples of coherent tidal streams are discovered, then
this would favor fluid dark matter models which produce smoother and highly 
spherical dark matter halos.

\section{Acknowledgements}

The simulations were performed on the ORIGIN 3800 at the CINECA Supercomputing
Center in Bologna and on an ALPHA workstation at the University of
Washington. L.M. thanks Volker Springel for providing the code used to
construct the galaxy models and Elena d'Onghia for providing the algorithm
to fit King profiles to the simulations. L.M. was supported by the 
National Science Foundation (NSF Grant 9973209).

\end{document}